\documentclass[12pt]{article}
\usepackage{graphicx}

\newcommand\pubnumber{}
\newcommand\pubdate{}

\def\napoli{Institute for Nuclear Physics\\
Johannes Gutenberg-University Mainz, 55128 Mainz, GERMANY}
\def\support{}

\def\Title#1{\begin{center} {\Large #1 } \end{center}}
\def\Author#1{\begin{center}{ \sc #1} \end{center}}
\def\Address#1{\begin{center}{ \it #1} \end{center}}

\newcommand\pubblock{\rightline{\begin{tabular}{l} \pubnumber\\
         \pubdate  \end{tabular}}}
\newenvironment{Abstract}{\begin{quotation}  }{\end{quotation}}
\newenvironment{Presented}{\begin{quotation} \begin{center} 
             PRESENTED AT\end{center}\bigskip 
      \begin{center}\begin{large}}{\end{large}\end{center} \end{quotation}}

\def\beq{\begin{equation}}
\def\eeq#1{\label{#1}\end{equation}}
\def\eeqn{\end{equation}}

\def\beqa{\begin{eqnarray}}
\def\eeqa#1{\label{#1}\end{eqnarray}}
\def\eeqan{\end{eqnarray}}

\let\bar=\overbar

\def\Dslash{\not{\hbox{\kern-4pt $D$}}}
\def\dslash{\not{\hbox{\kern-2pt $\del$}}}

\def\msb{{\bar{\ssstyle M \kern -1pt S}}}

\begin{document}
\begin{titlepage}
\pubblock

\vfill
\Title{Measurement of meson transition form factors at BESIII}
\vfill
\Author{ Christoph Florian Redmer for the BESIII Collaboration\support}
\Address{\napoli}
\vfill
\begin{Abstract}
Meson transition form factors describe the coupling of photons and hadrons. They are an important input to the 
calculations of the light-by-light scattering contribution of the anomalous magnetic moment of the muon. At the BESIII 
experiment in Beijing, the transition form factors of pseudoscalar mesons are studied in their Dalitz decays, in their 
radiative production in $e^+e^-$ annihilation, and in their production in two-photon scattering. All relevant kinematic 
regimes are covered. An overview of the recent results and the status of the ongoing analyses is provided.
\end{Abstract}
\vfill
\begin{Presented}
Thirteenth International Conference on the Intersections of Particle and Nuclear Physics (CIPANP2018)\\
Palm Springs, California, May 28--June 3, 2018
\end{Presented}
\vfill
\end{titlepage}
\def\thefootnote{\fnsymbol{footnote}}
\setcounter{footnote}{0}

\section{Introduction}
Meson transition form factors (TFF) describe the interaction of hadronic matter with two photons of arbitrary 
virtuality. They represent the difference to point-like interactions, as e.g. described by QED. Thus, TFFs also contain 
valuable information on the structure of the mesons. Additional interest in TFF comes from the anomalous magnetic moment 
of the muons $a_\mu$. Here, TFF are needed as experimental input for the Standard Model (SM) prediction.

Being one of the most precisely determined parameters in the SM, $a_\mu$ is known to $0.5\,\textrm{ppm}$ in experiment 
as well as theory. However, there is a long standing discrepancy of more than three standard deviations between the 
direct measurement and the SM prediction~\cite{JegNyf}. Since the deviation might be a hints for New Physics, numerous 
activities in experiment and theory have been initiated in order to test if the discrepancy is 
significant~\cite{Blum:2013xva}.

To date, the most precise direct measurement of $a_\mu$ comes from the BNL experiment E821~\cite{Bennett:2006fi}. Two 
new experiments are being prepared, which aim at direct measurement with a fourfold increase of the accuracy. On the 
one hand, the E989 experiment at Fermilab~\cite{Grange:2015fou} reuses the BNL storage ring and tries to reduce the 
uncertainties with higher beam intensities and an improved apparatus. On the other hand, the new experiment at 
J-PARC~\cite{Mibe:2011zz} uses a beam of ultra cold muons, which allows to avoid the use of focusing electric fields. As 
a consequence, there will be two systematically independent direct measurements of $a_\mu$.

The SM prediction of $a_\mu$ contains contributions of all three fundamental interactions. While the final result is 
almost completely given by the contribution of QED, the uncertainty is completely dominated by the hadronic 
contributions. Due to the running of the strong coupling constant, these, in contrast to the QED and weak contributions, 
cannot be calculated using perturbative methods. The hadronic contributions are separated into two parts, the hadronic 
vacuum polarization contribution $a_\mu^{hVP}$, and the hadronic light-by-light scattering contribution $a_\mu^{hLbL}$. 
Recent activities in lattice QCD demonstrated the feasibility of determining the hadronic contributions on the lattice, 
however, the current accuracy is not yet sufficient to compete with the upcoming experimental accuracy~\cite{lqcd}. 
Another approach is to exploit experimental information to improve the calculations. There is a dispersive approach 
which allows to systematically improve the knowledge on $a_\mu^{hVP}$ by measuring hadronic cross sections at $e^+e^-$ 
machines~\cite{hvp}. The determination of $a_\mu^{hLbL}$ is more involved. In the past, hadronic models were used to 
evaluate the light-by-light contribution~\cite{JegNyf,Gla09}. All approaches agree on the importance of the individual 
subprocesses to the light-by-light scattering, as suggested by de~Rafael~\cite{sort}. The pole contributions of the 
pseudoscalar mesons $\pi^0, \eta$ and $\eta^\prime$ are most important, followed by the contributions of pion loops 
and heavier resonances. Recently, dispersive approaches have been designed~\cite{mainz,bern}, aiming at a model 
independent determination of the most important contributions, which also reduces the overall model dependence of 
$a_\mu^{hLbL}$. The most important experimental observables needed as input to these calculations are the pseudoscalar 
meson TFFs at arbitrary virtualities and the partial waves of the process $\gamma^{(*)}\gamma^{(*)}\to\pi\pi$.

Experimentally, TFFs can be determined in three different processes. In the time-like regime, TFFs are measured in 
Dalitz decays and in the radiative production of pseudoscalar mesons in $e^+e^-$ annihilation. Both cases allow 
to study the TFFs as function of a single virtuality, which corresponds to the squared mass of the lepton pair. In the 
former case, the mass can vary between the rest mass of the leptons and the rest mass of the decaying meson. In the 
latter case, the virtuality is fixed to the energy of the collider.

The space-like regime can be studied in $e^+e^-$ scattering processes, where each of the beam leptons emits a photon. 
The two photons fuse to form a hadronic state. In principle, this process allows to study meson TFFs at arbitrary 
space-like virtualities. Limitations are given by the detector setups. The smaller the virtualities of the photons, the 
smaller are the scattering angles of the leptons. The cross sections of two-photon scattering processes decrease with 
increasing virtualities. The investigation of two-photon scattering is typically divided into three categories, 
depending on the possibility to detect none, one, or both of the scattered leptons in the detector.

The BESIII detector is well suited to study meson TFFs in the time-like and space-like regime using all of the above 
approaches.

\section{The BESIII detector}
The BESIII detector is a magnetic spectrometer~\cite{Ablikim:2009aa} located at the Beijing Electron Positron Collider 
(BEPCII)~\cite{Yu:IPAC2016-TUYA01}. The cylindrical core of the BESIII detector consists of a helium-based  multilayer 
drift chamber (MDC), a plastic scintillator time-of-flight system (TOF), and a CsI(Tl) electromagnetic calorimeter 
(EMC), which are all enclosed in a superconducting solenoidal magnet providing a 1.0~T magnetic field. The solenoid is 
supported by an octagonal flux-return yoke with resistive plate counter muon identifier modules interleaved with steel. 
The acceptance of charged particles and photons is 93\% over $4\pi$ solid angle. The charged-particle momentum  
resolution at $1~{\rm GeV}/c$ is $0.5\%$, and the $dE/dx$ resolution is $6\%$ for the electrons from Bhabha scattering. 
The EMC measures photon energies with a resolution of $2.5\%$ ($5\%$) at $1$~GeV in the barrel (end cap) region. The 
time resolution of the TOF barrel part is 68~ps, while that of the end cap part is 110~ps.

The accelerator BEPCII provides $e^+e^-$ collisions at center-of-mass energies between $\sqrt{s}=2.0\,GeV$ and
4.6\,GeV. The luminosity is optimized for data taking at the peak of the $\psi(3770)$ resonance, i.e. $\sqrt{s} = 
3.773\,\textrm{GeV}$. The design luminosity of $10^{33}\,\textrm{cm}^{-2}\textrm{s}^{-1}$ has been reached. Over the 
past years large data samples have been collected at the $J/\psi$ resonance~\cite{Ablikim:2016fal}, $\psi(3686)$ 
resonance~\cite{Ablikim:2017wyh}, as well as at and above the $\psi(3770)$ 
resonance~\cite{Ablikim:2013,Ablikim:2015nan}, which are used to pursue the BESIII physics program, focusing on charm 
physics, charmonium and charmoniumlike spectroscopy, light hadron physics, QCD tests, and precise $\tau$ mass 
measurements.

\section{Dalitz decays of light mesons}
A recent example of the investigation of Dalitz decays of pseudoscalar mesons at BESIII, is the measurement of the 
decay $\eta^\prime\to\gamma e^+e^-$~\cite{Ablikim:etap}. It is the first measurement of the $\eta^\prime$ Dalitz decay 
with an $e^+e^-$ pair in the final state. The mesons are tagged by the monochromatic photon in the radiative decay of 
the $J/\psi$ resonance. From the data set of $1.31\cdot10^9$ inclusive $J/\psi$ decays, $864\pm36$ events with a Dalitz 
decay of the $\eta^\prime$ are reconstructed. The most severe background contribution is due to photon conversion in the 
beam pipe and the detector material. By tracking the lepton pairs back to a common vertex, which is required to be less 
than 2\,cm away from the interaction point in radial direction, this background could be sufficiently suppressed. The 
branching ratio is determined as $\mathcal{B}(\eta^\prime\to\gamma e^+e^-) = 4.69\pm0.20_{\rm stat}\pm0.23_{\rm syst}$.

\begin{figure}
 \begin{minipage}{0.48\textwidth}
  \includegraphics[height=5cm]{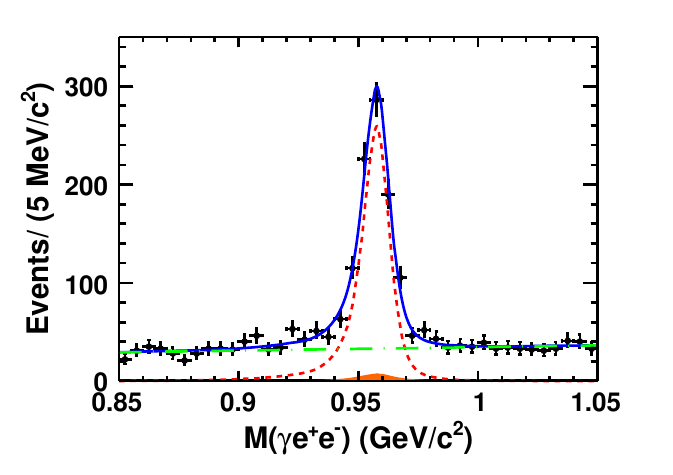}
 \end{minipage}\hfill%
 \begin{minipage}{0.48\textwidth}
  \includegraphics[trim=0mm 0mm 0mm 1mm, clip, height=5cm]{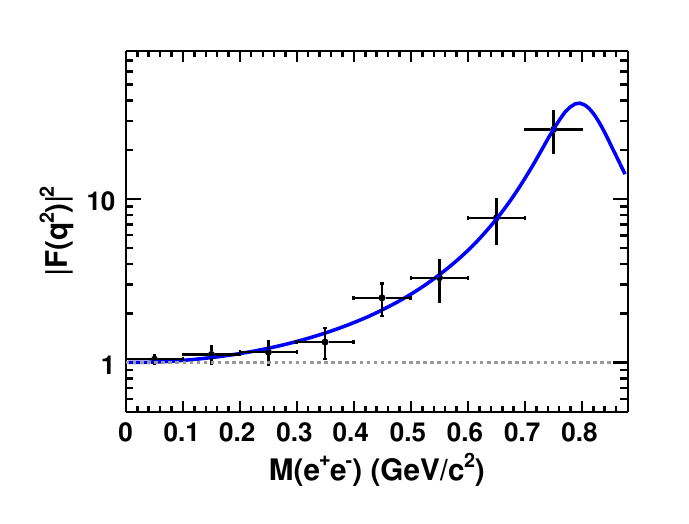}
 \end{minipage}
 \caption{\label{fig:etap}(From Ref.~\cite{Ablikim:etap}) \textbf{left:} Invariant mass of $e^+e^-\gamma$. The dashed 
green line shows the background contribution from the fit result. The solid red histogram sows the remaining background 
form photon conversion. \textbf{right:} $\eta^\prime$ TFF as function of the invariant mass of $e^+e^-$ and fitted with 
Eq.\ref{eq:epsp} (blue line).}
\end{figure}

The mass of the lepton pair is fitted with a single pole approximation of the TFF
\begin{equation}\label{eq:epsp}
 |F(q^2)|^2_{\eta^\prime} = \frac{\Lambda^2(\Lambda^2-\gamma^2)}{(\Lambda^2-q^2)^2-\Lambda^2\gamma^2}, 
\end{equation}
where $\Lambda$ and $\gamma$ correspond to the mass and with of an effective vector meson contributing to the decay. 
The fit results in $\Lambda = (0.79\pm0.04_{\rm stat}\pm0.02_{\rm syst})\,\textrm{GeV} $ and $\gamma = 
(0.13\pm0.06_{\rm stat}\pm0.03_{\rm syst})\,\textrm{GeV}$. The slope parameter of the TFF is determined as 
$1.60\pm0.17_{\rm stat}\pm0.08_{\rm syst}$, which is in good agreement with the Lepton-G result, based on the Dalitz 
decay with a muon pair, and the extrapolation of the space-like CELLO result to the time-like region. The accuracy is 
at the level of the space-like extrapolation, which is an improvement compared to the previous time-like result.

\section{Radiative production of light mesons}
The study of the radiative decays $\psi(3686)\to\gamma\pi^0,\eta\eta^\prime$ is a recent result on the radiative  
production of pseudoscalar mesons~\cite{Ablikim:radpsip}. The motivation of this analysis also comes form the need to 
improve the understanding of radiative transitions of charmonium resonances. A data set of $448\cdot10^6$ inclusive 
$\psi(3686)$ decays is analyzed. The pseudoscalar mesons are reconstructed from their most abundant decay modes, where 
similarities in the topologies between the final states are exploited: $\pi^0\to\gamma\gamma, \eta\to\pi\pi\pi$ and 
$\eta^\prime\to\eta\pi\pi$, with charged and neutral pions in the $\eta/\eta^\prime$ decays, respectively. The fully 
neutral final state of $\psi(3686)\to\gamma\pi^0$ requires additional constraints to suppress background from photon 
conversion. The most severe background contribution comes from $e^+e^-\to\gamma\gamma$. It can be suppressed by counting 
the single hits in the MDC in the azimuthal section between the two photon candidates, which are assigned to the 
$\pi^0$. In total $56053.5\pm980.8$ events are found for the radiative decay to $\eta^\prime$, $382.5\pm78.9$ for the 
case of $\eta$, and $423.4\pm71.4$ for $\psi(3686)\to\pi^0\gamma$. In the latter two cases the numbers correspond to 
significances of $7.3\sigma$ and $6.7 \sigma$, respectively. The branching ratios are determined as 
$\mathcal{B}(\psi(3686)\to\eta^\prime\gamma) = 125.2\pm2.2_{\rm stat}\pm6.2_{\rm syst}$, 
$\mathcal{B}(\psi(3686)\to\eta\gamma) = 0.85\pm0.18_{\rm stat}\pm0.04_{\rm syst}$, and 
$\mathcal{B}(\psi(3686)\to\pi^0\gamma) = 0.95\pm0.16_{\rm stat}\pm0.05_{\rm syst}$.

\begin{figure}
 \includegraphics[width=\textwidth]{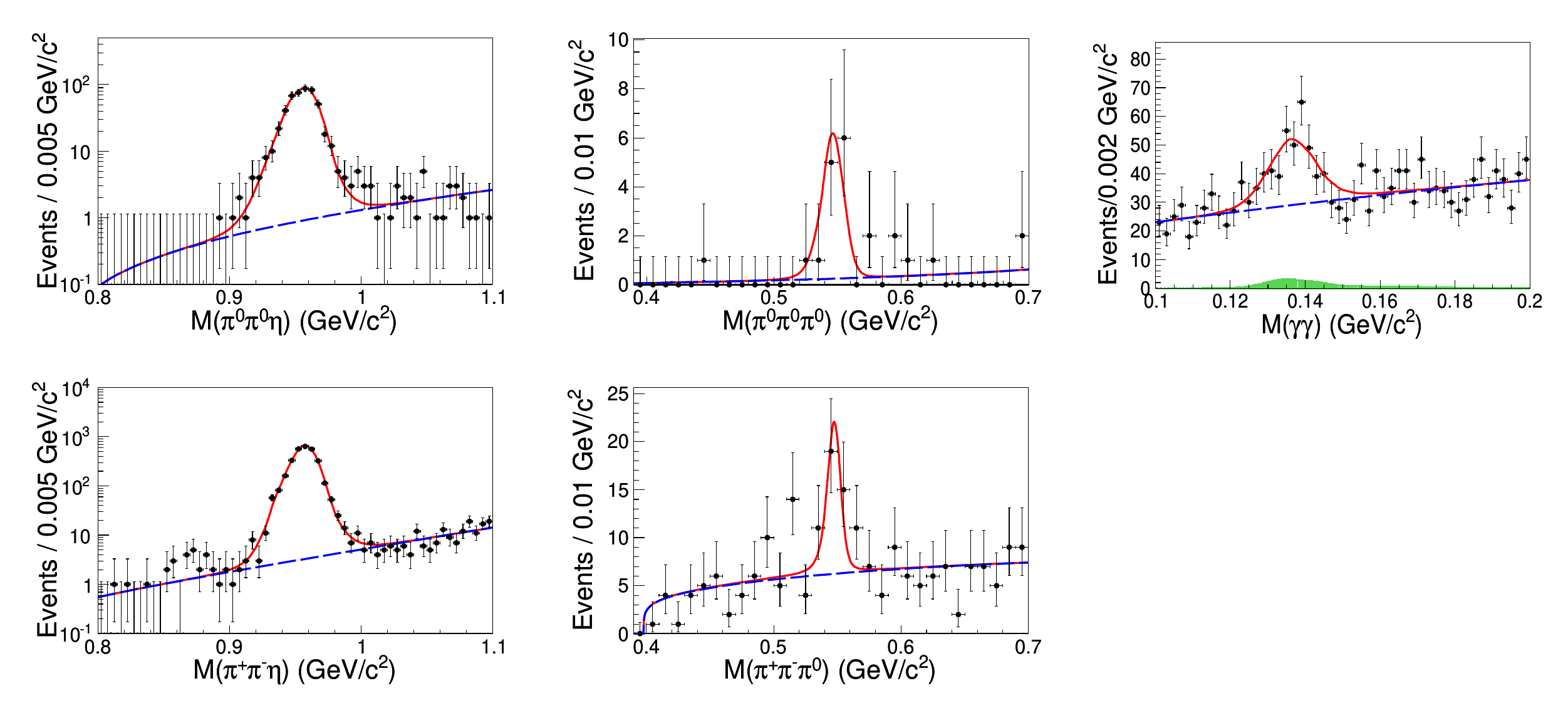}
 \caption{\label{fig:psip} (From Ref.~\cite{Ablikim:radpsip}) Mass distributions of individual final states. Left to 
right: $\psi(3686)\to\gamma\eta^\prime, \psi(3686)\to\gamma\eta$ and $\psi(3686)\to\gamma\pi^0$. Top(Bottom): Neutral 
(Charged) decay modes. The red line illustrates the total fit result. The blue line is the continuum background. The 
green histogram shows the contribution of $\chi_{cJ}\to\pi^0\pi^0$.} 
\end{figure}

The analysis is currently extended to the continuum region. The process $e^+e^-\to\gamma \pi^0, \eta, \eta^\prime$ can 
be used to study the momentum dependence of TFFs in the range $4.0\leq q^2\,[\textrm{GeV}^2]\leq21.16$. The pQCD 
prediction that time- and space-like TFFs behave identical for large momentum transfers $q^2$ can be tested. If this is 
already satisfied in the energy range covered at BESIII, it also allows to shed light on the ``BaBar-Belle 
puzzle''~\cite{Babar09,Belle12}, where the space-like TFFs deviate for $q^2<-10\,\textrm{GeV}^2$.

\section{Two-photon physics}
Compared to the annihilation cross sections, two-photon scattering is a rare process. Thus, only the largest data sets 
are used to study space-like TFFs of pseudoscalar mesons at BESIII. The momentum dependence of the TFF is studied in a 
single-tag technique, where only one lepton is scattered into the detector and can be measured along with the decay 
products of the produced meson. Using energy and momentum conservation, the momentum of the unmeasured lepton can be 
reconstructed. By placing a corresponding condition on the polar angle, a minimum momentum transfer, and thus the 
exchange of a quasi-real photon, is required. The same method was applied by CELLO~\cite{Cello91}, CLEO~\cite{Cleo98}, 
BaBar~\cite{Babar09}, and Belle~\cite{Belle12}, however, in the momentum range of $Q^2 = -q^2 \leq 1\,\textrm{GeV}^2$, 
which is of special importance for $a_\mu^{hLbL}$~\cite{Nyffeler:2016gnb}, the available information is scarce.

The investigation of the $\pi^0$ TFF selects event candidates with one lepton and at least two photons. The data set of 
$2.93\,\textrm{fb}^{-1}$ at $\sqrt{s}=3.773\,\textrm{GeV}$ is evaluated. An overwhelming background contribution of 
radiative Bhabha scattering is effectively suppressed by requiring the polar angle of the unmeasured lepton to be 
$|\cos\theta_{\rm miss}| \leq 0.99$ and the helicity angle of the photons from the $\pi^0$ decay to be $|\cos\theta_{\rm 
H}| \leq 0.8$. Additional background from incompletely reconstructed hadronic final states is rejected using the 
observable $R$, based on energy and momentum conservation, which was introduced by BaBar~\cite{Babar09} to suppress 
radiative effects in two-photon scattering.

\begin{figure}
 \centerline{\includegraphics[width=0.5\textwidth]{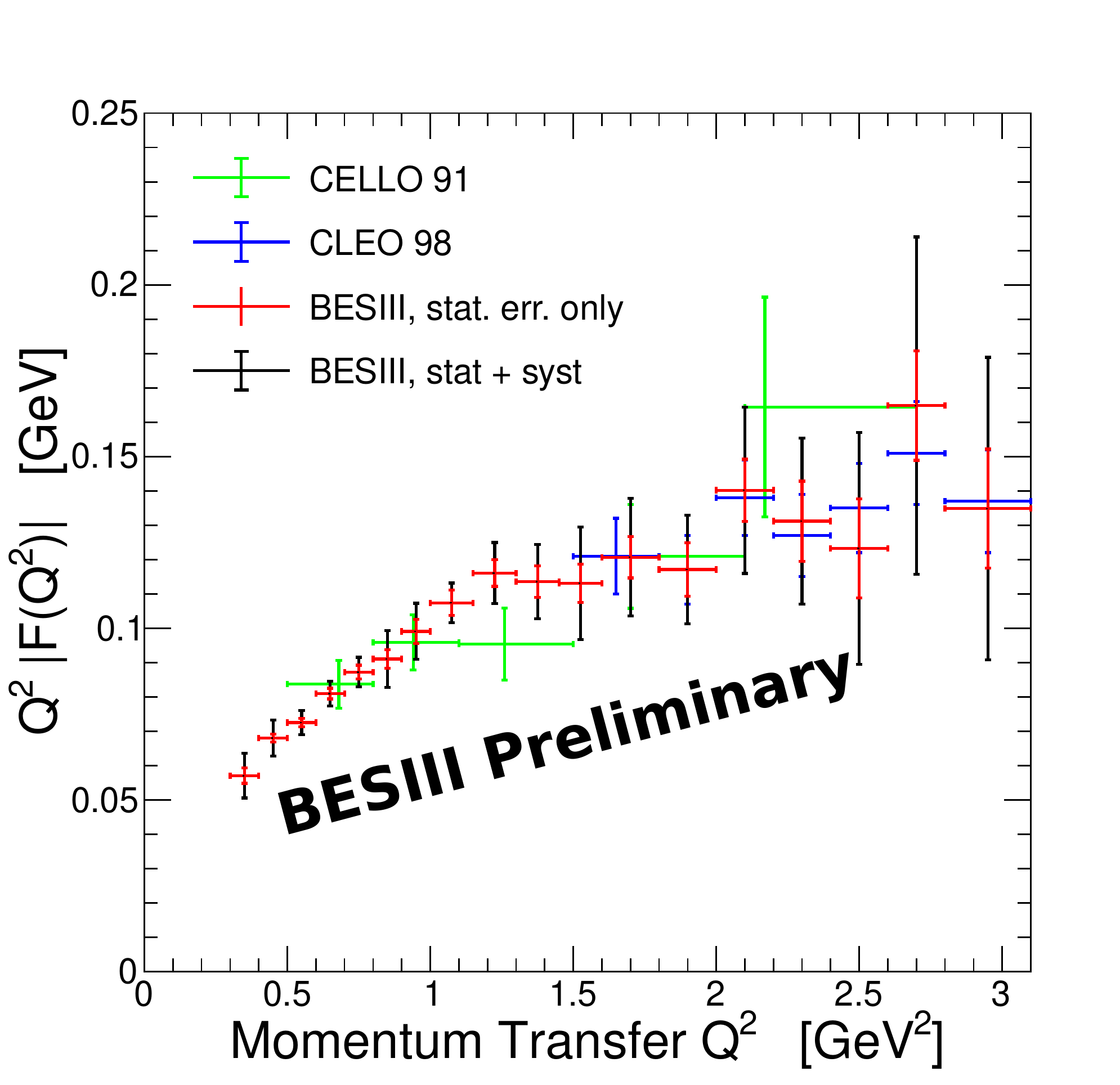}}
 \caption{\label{fig:pi0} Preliminary result of the $\pi^0$ TFF from BESIII (red: statistical uncertainty; black: 
total uncertainty). The results of the CELLO~\cite{Cello91} and CLEO~\cite{Cleo98} collaborations are show for 
comparison.}
\end{figure}

Remaining background is subtracted in a data-driven approach. In every bin of momentum transfer $Q^2$ the number of 
$\pi^0$ signals in the invariant mass of the decay photon candidates is fitted. By normalizing the event yield to the 
reconstruction efficiency and the integrated luminosity, the differential cross section is calculated. The TFF as a 
function of $Q^2$ is obtained by dividing out the point-like cross section. Figure~\ref{fig:pi0} shows the preliminary 
result of the measurement displayed as $Q^2\cdot|F(Q^2)|$. Statistical errors as well as the total uncertainty estimate 
are shown, while the latter does not yet contain the uncertainties due to radiative effects. These are under evaluation 
based on the recently released \textsc{Ekhara3.0}~\cite{Czyz:2018jpp}.

The BESIII result covers momentum transfers between $0.3\leq Q^2\,[\textrm{GeV}^2]\leq3.1$. It can be seen from 
Fig.~\ref{fig:pi0} that information on the $\pi^0$ TFF is provided with unprecedented accuracy for 
$Q^2\leq1.5\,\textrm{GeV}^2$, which is the region most relevant for $a_\mu^{hLbL}$. For larger values of $Q^2$ the 
accuracy is still compatible with previous measurements.

An analysis of the TFF of the $\eta$ and $\eta^\prime$ mesons is performed analogously. Currently, only the charged 
decay modes $\eta\to\pi^+\pi^-\pi^0$ and $\eta^\prime\to\pi^+\pi^-\eta$ are considered, allowing for a common analysis 
strategy looking for the decay of $\pi^0/\eta$ in two photons. Based on the data taken at $\sqrt{s}=3.773\,\textrm{GeV}$ 
a competitive statistical accuracy in the momentum range $0.3\leq Q^2\,[\textrm{GeV}^2] \leq3.1$ is achieved compared 
to previous measurements, which combined several decay modes~\cite{Cello91,Cleo98}.

Another important aspect of the two-photon physics program at BESIII is the measurement of the reaction 
$\gamma\gamma^\ast\to\pi^+\pi^-$. So far, only the production by two quasi real photons was studied~\cite{twopi}.
The analysis at BESIII follows the single-tag strategy, successfully applied to investigate the production of single 
pseudoscalar mesons. The main background contributions stem from the two-photon production of muon pairs, and the 
radiative Bhabha scattering process, in which the photon couples to a $\rho$-meson, decaying into two pions. While the 
first kind of background contribution can be efficiently rejected by combining different detector information for 
particle identification in a boosted decision tree, which can be trained by established MC generators~\cite{MuGen}, 
the latter kind of background is irreducible. It is removed by fitting and subtracting the $\rho$ peak in the 
$\pi^+\pi^-$ invariant 
mass. The remaining events allow to study for the first time the reaction $\gamma\gamma^\ast\to\pi^+\pi^-$ at  momentum 
transfers $0.2\leq Q^2 [\textrm{GeV}^2] \leq2.0$, invariant masses between $2m_{\pi}\leq M_{\pi\pi}[\textrm{GeV}] 
\leq2.0$, at a full coverage of the pion helicity angle $\cos\theta^*$. An equivalent analysis of neutral meson 
systems, i.e. $\gamma\gamma^\ast\to\pi^0\pi^0$ and $\gamma\gamma^\ast\to\pi^0\eta$ is in preparation.

\section{Outlook}
At BESIII, meson TFFs are studied in different kinematic regions. Important input to the calculations of the hadronic 
light-by-light scattering contribution to $a_\mu$ is provided. Based on existing data, unprecedented accuracy is 
achieved. The continued data taking allows to further improve the accuracy and investigate more rare processes.

An important next step in the investigation of two-photon scattering is the measurement of doubly-virtual meson TFFs. 
Due to the small cross section, published information was not available until recently~\cite{BaBar:2018zpn}. The BESIII 
collaboration has collected more than $10\,\textrm{fb}^{-1}$ at $\sqrt{s}\geq3.773\,\textrm{GeV}$, which are being 
analyzed to extract the TFF of $\pi^0,\eta$ and $\eta^\prime$ around $(Q^2_1\approx1\,\textrm{GeV}^2, 
Q^2_2\approx1\,\textrm{GeV}^2)$. The statistics is even expected to be sufficient to test the $Q^2$ dependence of 
different hadronic models~\cite{Nyffeler:2016gnb}.

\end{document}